\documentclass[11pt]{article}
\usepackage{graphicx}
\usepackage{amssymb}
\usepackage{epstopdf}
\begin{document}

\qquad\qquad\qquad\qquad\qquad\qquad Studies in History and Philosophy of\newline  
\indent \qquad\qquad\qquad\qquad\qquad\qquad\thinspace\thinspace Modern Physics $\bf{36}$, 716-723 (2005)

\qquad\qquad\qquad\qquad\qquad\qquad\qquad\qquad  \LARGE{Book Review}

\normalsize{\textbf{Stephen L. Adler, Quantum theory as an emergent phenomenon, Cambridge University Press, Cambridge, ISBN 0521831946, 2004, 238pp.}}

Reviewed by Philip Pearle, Hamilton College, Clinton NY 13323 (ppearle@hamilton.edu).

	The enlightenment task, of trying to explain the unnatural by the natural---in this case, the ``unnatural"  being quantum physics and the ``natural" being classical physics---was begun soon after the codification of quantum theory, in 1926, by Louis deBroglie and Erwin Madelung.  These, and later approaches by David Bohm, Edward Nelson and others, could be regarded as half-measured.  Classical particles and their dynamics are re-introduced, but a strong element of the unnatural remains.
In the deBroglie--Bohm and Madelung models, it is the mysterious quantum force. In the Nelson model,   
it is the mysterious backward diffusion process (which, together with the usual classical forward diffusion process, forces a particle's drift---its mean position---to be a dynamically determined quantity instead of, as classically, an independent variable set by external influences).  

	Stephen Adler's extraordinary work is full-measured.  The goal is to obtain relativistic quantum field theory (and, incidentally, thereby, its slow-speed limit, non-relativistic quantum mechanics) as a statistical mechanics canonical ensemble average of classical variables obeying classical dynamics.  The argument is laid out very carefully, with honest and extensive discussion of requirements and assumptions and great attention to detail.  Reading the book, I was reminded of a detective-adventure  story, where the author initially, comfortingly, assures one of the safe outcome, but one has strong interest in the twists and turns of the story, and whether the author can truly resolve the dilemmas to one's satisfaction.  In usual reviews of detective-adventure stories it is forbidden to give away the plot, but for this review I believe it is mandatory.  Although that  has aspects of the worst kind of ``book report," I think it is the only way I can fulfill my obligation to the reader and convey the richness and some of the subtlety of this unique contribution.  
	
	The  basic classical dynamical  variables are $N\times N$ matrices $q_{r}$, $p_{r}$.  Those with complex elements are called bosonic variables, while those called fermionic variables have elements which are  Grassmann numbers (complex numbers multiplying anticommutating objects).  The physical assignment of these matrix variables is not made until much later so that the argument is always as general and unrestricted as it can be.  However, the naming of these variables prefigures their final identification, i.e., a $q_{r}$ will eventually be chosen to represent a field amplitude---more precisely, $N^{2}$ amplitudes---in a small volume surrounding the spatial point ${\bf x}_{r}$.    The choices are made that the bosonic variables are self-adjoint and, for fermionic variables, $p_{r}=q_{r}^{\dagger}$.   
	
	All important quantities of classical mechanics appear here.  The action, Lagrangian, Hamiltonian, generators of canonical transformations, are formed from the trace of products of the matrix variables and their sum with constant ({\it not} matrix) coefficients, which is why the author calls this ``Trace Dynamics."   The derivative of a trace quantity ${\bf A}\equiv$Tr$A$ with respect to a matrix variable is defined, e.g., $\delta {\bf A}/\delta q_{r}$ is a matrix,  an element of which is the derivative with respect to $q_{r}$'s transposed matrix element.  This allows definition of a (trace)  Poisson bracket,  $\{{\bf A}\,{\bf B}\}\equiv$ Tr$\sum_{r}\epsilon_{r}[\delta{\bf A}/\delta  q_{r}\delta{\bf B}/\delta p_{r}-\delta{\bf B}/\delta  q_{r}\delta{\bf A}/\delta p_{r}]$ ($\epsilon_{r}= 1$ or -1, depending upon whether the $r$ labeled variables are bosonic or fermionic), obeying the Jacobi identity. For an example of its useage, it is shown that $d{\bf A}/dt=\partial {\bf A}/\partial t+\{{\bf A}\,{\bf H}\}$:  if  one wants to extract from this the dynamical equation  $dp_{r}/dt=-\delta {\bf H}/\delta q_{r}$, one must apply it to ${\bf A}=$Tr$p_{r}j$  (where $j$ is an arbitrary constant matrix) rather than to  ${\bf A}={\bf p}_{r}$ which can only give the equation of motion for the diagonal elements of $p_{r}$.  
		
	There are three conserved quantities of special importance.  The first is the trace Hamiltonian {\bf H} itself (with $H$ assumed to be time-independent, self adjoint, and formed with as many fermionic $q$'s as $p$'s in its matrix products).  The Poisson bracket formalism is invariant under canonical transformations, and ${\bf H}$ is the infinitesimal canonical generator of time translations.  
The second, ${\bf N}\equiv i\sum_{F}q_{r}p_{r}$ (the sum is limited to fermionic variables), is evocatively called the ``trace fermion number," but, of course, these are classical fields which do not describe particles. ${\bf N}$ is the infinitesimal canonical generator of phase transformations of the fermionic variables, under which the Lagrangian is invariant.  	
	
	When a theory provides a ``click," something neat which drops out of the mathematics, one's attention perks up.  Such is the third conserved quantity---or, rather, $N^{2}$ conserved quantities---uncovered by Adler's student Andrew Millard, 
\[	\tilde C\equiv \sum_{B}[q_{r}, p_{r}] -\sum_{F}\{q_{r}, p_{r}\}
\]
\noindent  ([,] and \{,\} denote commutator and anticommutator).  
$\tilde C$ is traceless and anti-self-adjoint.  
That a commutator  should appear in a fundamental way in this classical theory is surprising, since the matrix dynamical variables can take on any  values and commutation plays no role.  The reason  behind these conserved quantities is that the Lagrangian, like any trace quantity, is invariant under unitary transformations of the $q$'s and $p$'s  (these are special cases of canonical transformations).  The $N^{2}$ conserved quantities making up $\tilde C$ arise from Noether's theorem applied to the $N^{2}$ independent unitary transformations.  It is a ``thermal" average of $\tilde C$ which will morph into $i\hbar$.  
	
	For, after dynamics, comes statistical mechanics. Introduced is the phase space measure $d\mu$ (the product of the independent real and imaginary parts of $d(q_{r})_{ij}$ and $d(p_{r})_{ij}$).  Liouville's theorem holds:  $d\mu$ is shown to be invariant under general canonical transformations and so, in particular, under time evolution.  The dynamics of the physical system is assumed to be complicated enough so that its microcanonical ensemble obeys the ergodic hypothesis, i.e., is uniformly spread out over the available phase space constrained by constant ${\bf H} $,  ${\bf N}$ and $\tilde C$.  The equilibrium probability density distribution $\rho$ of a subsystem is obtained in the usual canonical ensemble way, by maximizing the entropy subject to the constraints of constant ${\bf H} $,  ${\bf N}$ and $\tilde C$ expectation values. The result is  $\rho=Z^{-1}\exp-[\hbox{Tr}\tilde \lambda\tilde C+\tau {\bf H} +\eta {\bf N}]$, where the Lagrange multipliers $\tau$, $\eta$ and the traceless and anti-self-adjoint matrix elements $\tilde \lambda_{ij}$  are inverse ``temperatures"  (e.g., like the chemical potential, which is the inverse ``temperature" associated with conservation of particle number in the usual grand canonical ensemble).  
	
	The mean value of any polynomial in the dynamical variables, $A$, is denoted $\langle A\rangle_{\hbox {AV}} \equiv\int d\mu\rho A$.  A unitary transformation of this equation and use of the unitary invariance of  $d\mu$, ${\bf H} $ and ${\bf N}$ shows that {\it any} $\langle A\rangle_{\hbox {AV}} $ must be a function of  $\tilde \lambda$, the only non-dynamical matrix in $\rho$.  $\tilde \lambda$  can be expressed in any basis, so choose the basis  in which it is diagonal.   Choose  $A=\tilde C$: then, in this basis,  $\langle \tilde C\rangle_{\hbox {AV}}$ is also diagonal. 
	
	 Now, a crucial  assumption is made, that these diagonal elements of  $\langle \tilde C\rangle_{\hbox {AV}}$ all have the same magnitude.  (The author suggests that this democratic behavior could eventually be understood as arising from initial conditions and/or from a deeper understanding of the dynamics.) That magnitude will be identified with $\hbar$.   Then, since $\tilde C$ is traceless and anti-self-adjoint, 
$N$ must be even (hereby assumed) and $\langle \tilde C\rangle_{\hbox {AV}}\equiv i_{\hbox{eff}}\hbar$, where $i_{\hbox{eff}}$ is a diagonal matrix with $N/2$ elements $+i$ and $N/2$ elements $-i$.  The assumption is implemented by choosing the canonical ensemble so that $\tilde \lambda$ is restricted to equal  $i_{\hbox{eff}}$ multiplying a constant, the lone remaining inverse temperature associated with conservation of  $\tilde C$.  About the magnitude of $\hbar$, surprisingly little is said, except for the comment ``our approach implies that it has a dynamical origin."  Presumably that magnitude depends upon the value of the associated inverse temperature.  One might speculate that initial conditions and dynamics could determine that temperature in the same sense that they determine the $\approx 3^{\circ}$ cosmic radiation temperature.  

	All dynamical $N\times N$ matrices are then written, with the help of the Pauli matrix algebra, 
in $2\times 2$ block form, where the upper left $N/2\times N/2$ block has $i_{\hbox{eff}}=i$ and will be responsible for 	quantum theory.  The lower right $N/2\times N/2$ block has $i_{\hbox{eff}}=-i$ and will be the complex conjugate of quantum theory, a poor relative which has to come along for the ride.  These diagonal blocks, which commute with $i_{\hbox{eff}}$, are labeled with the subscript eff (presumably for effective), while the two off-diagonal blocks, which anticommute with $i_{\hbox{eff}}$,  are not effective for the quantum theory consideration.  Just as anticommutation of the Pauli matrix $\tau_{3}$ with any $2\times 2$ matrix removes the off-diagonal part of that matrix,  the anticommutator $\{A,i_{\hbox{eff}}\}=2i_{\hbox{eff}}A_{\hbox{eff}}$ projects onto the effective  sector.

	The canonical ensemble probability density is still invariant under the subgroup of unitary transformations $U_{\hbox{eff}}$ which commute with $i_{\hbox{eff}}$.  It is because the phase space measure $d\mu$ includes integrating over this subgroup that any 
$\langle A\rangle_{\hbox {AV}}$ is a function of $i_{\hbox{eff}}$, which makes its upper left block just a number. The author wishes to exhibit $\langle A\rangle_{\hbox {AV}}$'s matrix nature. This is accomplished by restricting the integral range of $d\mu$ to a measure $d\hat\mu$, which eliminates integrating over the subgroup.  This so-called ``global unitary fixing," just requires appropriately fixing the numerical values of the matrix elements of one pair of bosonic variables (value of a bosonic field and its conjugate at one point of space), and integrating over the rest of the measure.  Although the matrix  $\langle A\rangle_{\hat{\hbox {AV}}}$ now depends upon the fixing, $\langle \bf{A}\rangle_{\hat{\hbox {AV}}}$ does not: the trace average is the same as if the integral was over $d\mu$ instead of over $d\hat\mu$.  It is later shown that, what is physically important,  the ``emergent" quantum theory's matrix elements, can be expressed as trace quantities.  A fixing to exhibit a specific matrix form for $\langle A\rangle_{\hat{\hbox {AV}}}$ is likened to choosing a gauge in electrodynamics to exhibit a specific functional form for the vector potential, although what is physically  important, the electromagnetic field, is independent of the choice of gauge. 

	The next step is to consider the ensemble average of specific quantities by utilizing an analogy to the equipartition theorem.  In standard statistical mechanics, by considering $0=\int d\mu\partial(\rho p)/\partial p$ with $\rho=Z^{-1}\exp-\beta H$, carrying through the derivatives results in $0=1-2\beta\langle p^{2}/2m\rangle_{\hbox {AV}}$.  Here is considered 
\[0=\int d\hat\mu\delta(\rho A)/\delta x_{r},
\]
\noindent where 
$x_{r}$ is a $q_{r}$ or $p_{r}$.  (The author calls these equipartition equations   ``Ward identities," since the Ward identity in e.g., quantum electrodynamics, involves similar manipulations.) The choice  $A\equiv$Tr$\{\tilde C, i_{\hbox{eff}}\}W=$Tr$\tilde C\{i_{\hbox{eff}},W\}$ is made, with $W$ a polynomial in the dynamical variables.  The derivative of  $\rho$ brings down three terms, involving the derivative of 1) Tr$\tilde \lambda\tilde C$, 2) $\tau {\bf H}$  and 3) $\eta {\bf N}$, while the derivative of $A$ creates two terms, one involving the derivative of 4) $\tilde C$, the other of 5) $W$.  By anticommuting with $i_{\hbox{eff}}$, term 1) drops out, and the rest of the terms then involve only effective matrices.  The canonical ensemble average of the sum of the remaining four terms vanishes, but it is argued (by adding a term $-\sum$Tr$j_{r}x_{r}$ to the exponent of $\rho$, and making variations in the arbitrary functions $j_{r}$) that even more vanishes: the canonical ensemble average of the sum of these 4 terms sandwiched between {\it arbitrary} polynomials in the effective dynamical variables.  

	Now, as they say, the plot thickens.  Appropriate to a detective story, terms 2) and 3) are bumped off, leaving just terms 4) and 5).  Also,  important approximations are made in term 5).  
	
	Term 2) is  $\sim\tau \dot x_{r\hbox{eff}}$Tr$\tilde C_{\hbox{eff}}i_{\hbox{eff}}W_{\hbox{eff}}$, where $\dot x_{r\hbox{eff}}$ represents the effective part of the right hand side of the dynamical equations $\dot q_{r} =\epsilon_{r}\delta {\bf H}/\delta p_{r}$ and $\dot p_{r} =-\delta {\bf H}/\delta q_{r}$.  To make this vanish, it is assumed that $\tau^{-1}$ is the Planck temperature, which is so large that $\tau \dot x_{r\hbox{eff}}$ will be negligibly small for the relatively slow rates of change $\dot x_{r\hbox{eff}}$ which we see in physics accessible to us.  However, when there is such a high temperature, one expects that a variable's rate of change will be comparably fast, not slow. So, the assumption entails more.  It is supposed that the Hamiltonian is such that there is a ``mass hierarchy"  which, at the top end, is  governed by the Planck mass and Planck time, at the bottom end by our accessible physics, so that $ x_{r\hbox{eff}}= x_{r\hbox{eff}}^{\hbox{fast}}+x_{r\hbox{eff}}^{\hbox{slow}}$ (like Brownian motion's superposed slow diffusion and rapid jiggles). When the ensemble average is taken, in order to make the other part of term  2) negligible, the one involving 
$\tau \dot x_{r\hbox{eff}}^{\hbox{fast}}$, it is assumed that, in those regions of phase space where it is large,  $\tilde C_{\hbox{eff}}$ which it multiplies is negligibly small.  The plausibility of these assumptions is discussed.  

	The result is unusual.  Usually, it is the Hamiltonian which governs the canonical distribution, but here it is $\tilde C$  which dominates.  Something else is achieved.  One might have wondered how it is possible to have a Lorentz invariant theory if the system is in a thermal bath, since a thermal bath  exists in a preferred frame (in which the momentum and angular momentum vanish---it is suggested that this could be the local co-moving frame).  But here, the bath and the associated Hamiltonian, which is not a Lorentz invariant, are out of it: it is $\tilde C$ which reigns, and that  is Lorentz invariant. What is missing is the choice of Hamiltonian which entails this behavior.  The author says, ``this is a task for the future," a not unreasonable position, given the ever-present difficulty of finding the right Hamiltonian for particle physics.   

	Term 3),  $\sim\eta \dot x_{r\hbox{eff}}$Tr$\tilde C_{\hbox{eff}}i_{\hbox{eff}}W_{\hbox{eff}}$, holds only for fermionic variables $x_{r}$.  $\eta$ is chosen to vanish. It is pointed out that,  since $\rho$ does then not depend upon {\bf N}, this implies that the ensemble average of {\bf N} vanishes, since an exchange of the fermionic $p_{r}$'s and $q_{r}$'s in the ensemble average integral changes the sign of  {\bf N} but leaves $d\hat\mu$ and $\rho$ unaffected.  However, no fundamental reason for this choice is given.  It seemed to me that a reasonable argument  might be to make a parallel with the chemical potential in the usual grand canonical ensemble, which may be set equal to zero, reducing it to the canonical ensemble, if there is no particle exchange between the system and its bath.  If the dynamical variables of the subsystem here represent fields localized at points in a volume of space, and the bath is the fields in the surrounding volume, there is indeed no motion of variables from one volume to the other, and ${\bf N}$ may be fixed for the subsystem, allowing $\eta=0$.  
	
	Term 5) arises from Tr$\tilde C_{\hbox{eff}}\delta W$.  It is assumed that the subsystem under consideration is so large that fluctuations in $\tilde C_{\hbox{eff}}$ are small, so that  $\tilde C_{\hbox{eff}}$ may be replaced by $\langle \tilde C_{\hbox{eff}}\rangle_{\hbox {AV}}=i_{\hbox{eff}}\hbar$.  It is also assumed that, in the ensemble averages, $W_{\hbox{eff}}( x_{r})\approx W(x_{r\hbox{eff}})$ to great accuracy.  I thought this assumption not as obviously attainable as the rest because the product of an even number of  $x_{r}$'s has the ineffective off-diagonal blocks contributing to effective diagonal blocks. This assumption implies that, nonetheless, the off-diagonal  block contribution to the slow dynamics is negligibly small.  
	
	The result of these assumptions is the vanishing of the canonical ensemble average of 	
\begin{equation}\label{one}
i_{\hbox{eff}}[W_{\hbox{eff}}, x_{r\hbox{eff}}] \pm\hbar\frac{\delta W(x_{\hbox{eff}})}{\delta x'_{r\hbox{eff}}}
\end{equation}
\noindent sandwiched between arbitrary polynomials in the dynamical variables ($x'_{r}$ is the variable conjugate to $x_{r}$, and the sign is - for bosonic $x_{r}=q_{r}$ and + otherwise).  
Eq.(1) represents a ``derivation," if you will, of Dirac's ad hoc rule of equivalence of a quantum commutator bracket and a classical Poisson bracket.  

	With  $W_{\hbox{eff}}=H_{\hbox{eff}}$, Eq.(1) gives the (sandwiched, averaged) vanishing of 
\begin{equation}\label{two}
i_{\hbox{eff}}[H_{\hbox{eff}}, x_{r\hbox{eff}}] -\hbar\dot x_{r\hbox{eff}}.
\end{equation}	
\noindent If you were wondering how dynamics can come from averaging over a static canonical distribution, here's how.  Recall that $\dot x_{r}$ represents the function of $x$'s given by the Poisson bracket equation of motion.   That equation of motion determines the underlying dynamical solution $x_{r}(t)$, whose effective part can be taken at any time.  Eq.(2) says that $[x_{r\hbox{eff}}(t+\Delta t)-x_{r\hbox{eff}}(t)]/\Delta t$ is just what is given by the Heisenberg equation of motion in quantum theory.  Neat, huh?  (Appropriate to an adventure story, the line from the song in ``Casablanca" comes to my mind, ``The fundamental things apply as time goes by.")   From this, the (sandwiched, averaged) usual Heisenberg equation of motion for any polynomial of the $x$'s follows.

	With  $W_{\hbox{eff}}=$constant$\times x_{r\hbox{eff}}$ (the constant is a c-number for a bosonic variable, a Grassmann number for a fermionic variable), Eq.(1) yields the (sandwiched, averaged) canonical commutation (anticommutation) relations for bosonic (fermionic) variables.  
	
	With  $W_{\hbox{eff}}$ such that the infinitesimal generator of a canonical transformation is $\bf{W}$, Eq.(1) implies that this canonical transformation for the effective variables can be implemented by a unitary transformation with $W_{\hbox{eff}}$ as the infinitesimal generator.  
	
	It is now just a short hop to quantum field theory.  $H_{\hbox{eff}}$ is restricted to have a nondegenerate lowest eigenvalue with eigenvector $\psi_{0}$.  Then, the correspondence is made that the upper left block of 
\begin{equation}\label{three}
\psi_{0}^{\dagger}\langle A(x_{\hbox{eff}})\rangle_{\hat{\hbox {AV}}}\psi_{0}=\langle \hbox{vac}|A(X)|\hbox{vac}\rangle,
\end{equation}
\noindent where $A$ is any polynomial in $x_{r\hbox{eff}}$'s, interpreted (as previously mentioned)
as quantum fields.  The demonstrated properties of the left side of Eq.(3) imply that it behaves as a Wightman function, i.e.,  the right side of Eq.(3) with $X$'s as quantum fields, in terms of which local relativisitic quantum field theory can be expressed.  

	To show that the left side of Eq.(3) can be written as an ensemble average of a trace and so, as promised, is independent of the gauge-like ``fixing," it is noted that $\psi_{0}\psi_{0}^{\dagger}=(2\pi i)^{-1}\int dz (z-H_{\hbox{eff}})^{-1}$, where the integration is over an infinitesimal contour surrounding the lowest eigenvalue of  $H_{\hbox{eff}}$.  Then, the left side of Eq.(3) is $\langle$Tr$A(x_{\hbox{eff}})\psi_{0}\psi_{0}^{\dagger}\rangle_{\hat{\hbox {AV}}}$.  
	
		It is now straightforward to go from the Heisenberg picture just obtained to the Schr\"odinger picture and Schr\"odinger's equation, utilizing the demonstrated equivalence of the canonical time translation generated by $\bf{H}$, and the unitary time translation generated by $H_{\hbox{eff}}$.
		
		This is how the algebra of quantum theory ``emerges" from the statistical mechanics of the classical trace dynamics.  However, there is still one crucial thing missing from giving quantum theory entire: the Born probability interpretation.  
		
	This is achieved by tapping into the lexicon of dynamical wavefunction collapse models first proposed by this reviewer in the late 1970's, and well developed since.  In these models, the Schr\"odinger equation is altered by adding terms to the Hamiltonian, most importantly,  an anti-unitary operator multiplied by a randomly fluctuating function of white noise type (or a set of such mutually commuting terms). This makes a superposition of states (eigenstates of the anti-unitary operator or operators) dynamically evolve to one of the states.   Which state survives depends upon the particular white noise function. The probability associated with a particular final state is that associated to all the white noise functions which cause it, and is equal to the Born probability (i.e., to the squared magnitude of the state's coefficient in the initial superposition).  
	
	To achieve a collapse dynamics, fluctuations of  $\tilde C$ are taken into account. The assumption that $\tilde C$ may be replaced by
$\langle\tilde C\rangle _{\hat{\hbox {AV}}}=i_{\hbox{eff}}\hbar$ in term 5) of the ``Ward identity" is replaced by 
 $\tilde C=i_{\hbox{eff}}\hbar$ plus a ``fast" fluctuation,  the latter assumed to be well approximated by white noise behavior.  However, it is necessary that this fluctuation be self-adjoint if stochastic collapse dynamics is to be obtained.  It is pointed out that the anti-self-adjointness of $\tilde C$ is a consequence of the choice $p_{r}=q_{r}^{\dagger}$ for fermionic variables.  The more general choice of  $p_{r}=\sum_{s} q_{s}^{\dagger}A_{sr}$, with each $A_{sr}$ an $N\times N$ matrix (and $A^{\dagger}_{sr}=A_{rs}$) allows $\tilde C$ to have a self-adjoint part.  So, the substitution 
  $\tilde C=i_{\hbox{eff}}\hbar[1+{\cal K}(t)+{\cal N}(t)]$ is made in term 5) of the ``Ward identity," with ${\cal K}(t)$ a c-number and 
${\cal N}(t)$ a matrix. 

	This adds  ${\cal K}(t)$  and ${\cal N}(t)$-dependent terms to Eqs.(1) and (2).  Specializing to the upper left block where $i_{\hbox{eff}}=i$, the nonrelativisitic limit is taken, with the fermionic 
$x_{r}$'s replaced by quantum fields $\Psi_{r\hbox{eff}}$,  $\Psi^{\dagger}_{r\hbox{eff}}$  which annihilate and create a fermion at location $r$.  The modified Eqs.(2) for $\dot\Psi_{r\hbox{eff}}$ and  $\dot\Psi^{\dagger}_{r\hbox{eff}}$,  acting on the vacuum state, are taken to be the (unsandwiched,  unaveraged) field theory equations of motion, corrected for when there are fluctuations in $ \tilde C$.  The reason for the ``acting on the vacuum state" proviso is that when $i{\cal K}(t)$ has a real part and $i{\cal N}(t)$ has a self-adjoint part, necessary if these are to be white-noise dependent terms responsible for collapse, then the equation for $\dot\Psi^{\dagger}_{r\hbox{eff}}$ is not the adjoint of the equation for $\dot\Psi_{r\hbox{eff}}$.  

The equation for  $\dot\Psi_{r\hbox{eff}}$ acting on $|\hbox{vac}\rangle$ vanishes identically.  $\dot\Psi^{\dagger}_{r\hbox{eff}}|\hbox{vac}\rangle$'s equation of motion is converted to an equation of motion for the Schr\"odinger picture statevector $|\Phi,t\rangle$, which is a superposition of products of $\Psi^{\dagger}_{r\hbox{eff}}$'s acting on $|\hbox{vac}\rangle$.  This conversion requires the assumption that ${\cal N}(t)$ is the sum of operators each associated with a single $r$, i.e., ${\cal N}(t)=\sum_{r}{\cal N}_{r}(t)$ (${\cal K}(t)$ is specialized so that ${\cal N}_{r}(t)|\hbox{vac}\rangle=0$).  Neglecting the (non-collapsing) effect of real ${\cal K}(t)$  
and self-adjoint ${\cal N}(t)$, with ${\cal K}_{1}(t)\equiv$Im${\cal K}(t)$ and ${\cal M}_{1}(t)\equiv
(i/2)\sum_{r}m_{r}[{\cal N}_{r}(t)-{\cal N}^{\dagger}_{r}(t)]$, the statevector satisfies
\begin{equation}\label{four}
d|\Phi,t\rangle/dt=\hbar^{-1}[-iH_{\hbox{eff}}-{\cal K}_{1}(t)H_{\hbox{eff}}-(1/2){\cal M}_{1}(t)]|\Phi,t\rangle
\end{equation}
Eq.(4) does not yield  $d\langle\Phi,t |\Phi,t\rangle/dt=0$, but it is shown that this should be true for the choice of the more general $\tilde C$.  This discrepancy is attributed to neglect of nonlinear terms in $|\Phi,t\rangle$ when arriving at Eq.(4).  With addition of such terms to give constant $\langle\Phi,t |\Phi,t\rangle$, and the assumption that resulting collapse theory's density matrix evolution equation is of the Lindblad-Kossakowski type (so, for example, there is no superluminal signalling), the additional terms are uniquely determined.  

The  ${\cal K}_{1}(t)$ term in Eq.(4) yields collapse to $H_{\hbox{eff}}$'s eigenstates.  Collapse to energy eigenstates, first proposed by Donald Bedford and Derek Wang in the mid 1970's, has since then 
been advocated in formulations  by Gerald Milburn, Ian Percival, Daniel Fivel and most strongly perhaps by Lane Hughston, and has been explored by Stephen Adler and colleagues, with resulting experimental limits on the size of this term summarized in this book.  I have recently given a number of arguments that energy-driven collapse cannot give the behavior appropriate for a collapse model capable of describing a macroscopic world consistent with our experience.  For the simplest of examples, an isolated macroscopic object in a superposition of two locations will remain in such a superposition since the two states have precisely  the same energy spectrum (it is differences in energy spectra that allows one state to be selected as the end product of collapse over another ).  Adler agrees with this position, although the ${\cal K}_{1}(t)$ term may be there.  

So, he turns to discuss the ${\cal M}_{1}(t)$ mass-density proportional term.  It  is not precluded from being specialized to the form utilized by my 1989 Continuous Spontaneous Localization (CSL) collapse model, based upon combining my previous work with aspects of  GianCarlo Ghirardi, Alberto Rimini, and Tullio Weber's 1986 Spontaneous Localization (SL) model.  CSL does provide a satisfactory description of  the macroscopic world, with experimental outcomes occurring as predicted by the Born probability rule.  In this way, the theory presented  here can ``explain" both the structure of quantum theory and its interpretation.  

There you have it, every vicissitude overcome, every barrier gone over as well as either through or around, with impressive ingenuity, range and resourcefulness.  Is this the long sought formulation which makes quantum theory understandable? I'd say a definite ``maybe" or, to paraphrase Dr. Seuss in his moral tale of Horton, ``it could be, it could be, it could be like that." The author clearly was struck by the striking features of trace dynamics and its potential for attaining the ``emergent" quantum theory grail.  The argument satisfyingly flows along,  conveying that conviction, occasionally with reservations, to the reader.  The whole thing rests on the properties of an unknown Hamiltonian.  The assumptions about its behavior pile up, but they almost always appear natural, especially because, for the most part, they are exhaustively and evenhandedly explored so one is led to an appreciation of their necessity and even inevitability if the scheme is going to work.  Lacking the right Hamiltonian, one might nevertheless hope that the future can bring pursuit of ``toy models" to elucidate and confirm one or another feature of this remarkable construction.  

In his contribution to Paul Schilpp's collection of essays, ``Albert Einstein: Philosopher-Scientist," Einstein wrote his essay in 1949 at the Princeton Institute for Advanced Study, repeatedly referring to the ``statistical quantum theory."  He predicts that it ``would, within the framework of future physics, take an approximately analogous position to the statistical mechanics within the framework of classical mechanics."  How apt it is in this year of homage to Einstein to consider that the Albert Einstein Professor at the Princeton Institute for Advanced Study for 24 years, Stephen Adler, has produced just such a structure.

 \end{document}